\documentclass[amsmath,amssymb,nofootinbib]{revtex4}

\usepackage{latexsym,comment}
\usepackage{amssymb}
\usepackage{amsfonts}
\usepackage{amsmath,color}
\usepackage[dvips]{graphicx}
\usepackage{bm}

\def\mb#1{\mathbf{#1}}

\def\ber{\begin{eqnarray}}
\def\eer{\end{eqnarray}}
\def\beq{\begin{equation}}
\def\eeq{\end{equation}}

\begin{document}

\title{A Note on the Sagnac Effect and Current Terrestrial Experiments}

\author{Matteo Luca Ruggiero}
\email{matteo.ruggiero@polito.it}
 \affiliation{DISAT, Politecnico di Torino, Corso Duca degli Abruzzi 24, Torino, Italy\\
 INFN, Sezione di Torino, Via Pietro Giuria 1, Torino, Italy}
\author{Angelo Tartaglia}
\email{angelo.tartaglia@polito.it}
 \affiliation{DISAT, Politecnico di Torino, Corso Duca degli Abruzzi 24, Torino, Italy\\
 INFN, Sezione di Torino, Via Pietro Giuria 1, Torino, Italy}

\date{\today}

\begin{abstract}
We focus on the Sagnac effect for light beams in order to evaluate  if the higher order relativistic corrections of kinematic origin could be relevant for actual terrestrial experiments. Moreover, we discuss to what extent the analogy with the Aharonov-Bohm effect holds true in a fully relativistic framework. {We show that the analogy with the Aharonov-Bohm is not true in general, but is recovered in a suitable low order approximation, and that  even though the Sagnac effect is influenced by both  the position of the interferometer in the rotating frame and its extension, these effects are negligible for current terrestrial experiments.}
\end{abstract}

\maketitle

\section{Introduction}\label{sec:sdec}

More than a century ago, Sagnac first predicted and then verified that there is a shift of the interference pattern when an interferometer is set into rotation, with respect to what is observed when the device is at rest \cite{sagnac05,sagnac13}. Actually, if  $\bm{\Omega }$ is the (constant) rotation rate of the interferometer with respect to an inertial  frame, $\mathbf{S}$ is the vector associated to the area enclosed by the light path and $\lambda $ is the wavelength of light, he predicted, and actually measured, the following fringe shift (which was named, after him, ``Sagnac effect'')
\begin{equation}
\Delta z=4\frac{\mathbf{\bm{\Omega} \cdot S}}{\lambda c}
\label{eq:sagnac1}
\end{equation}
for  light waves counter-propagating in a rotating interferometer. The proper time difference associated to the fringe shift (\ref{eq:sagnac1}) turns out to be
\begin{equation}
\Delta t=\frac{\lambda }{c}\Delta z=4\frac{\mathbf{\bm{\Omega} \cdot
S}}{c^{2}} \label{eq:sagnac2}
\end{equation}
Actually, it is interesting to point out that Sagnac  interpreted these results in support of the ether theory against the
Special Theory of Relativity (SRT), since he wrote that   \textit{``[...] the observed interference effect turns out to be the optical vortex effect due to the motion of the system with respect to the ether [...]''}\cite{sagnac13}.  As a matter of fact, during the following years and also more recently (see e.g. the review \cite{RRinRRF} and the other papers  on the subject in the monograph \cite{RRF}), many authors interpreted this effect as a conundrum in the SRT when it is applied to rotating reference frames; however, it is nowadays manifest that the Sagnac effect can be completely explained in SRT as an observable consequence of   {the non-isotropy of the coordinate velocity of light,  related to the synchronization gap along a closed path in non-time-orthogonal frames (see e.g. \cite{found}, \cite{RRinRRF}, \cite{rizzi03a}, \cite{rizzi03b}, \cite{LL}, and references therein)}. Besides being relevant for the understanding of the theoretical foundations of the theory of relativity, the Sagnac effect has become important since the development of lasers, which allowed a remarkable advance of light interferometry\cite{post67}.

Thanks to the increasing accuracy, many technological applications based on the Sagnac effect were developed, such as fiber optic gyroscopes, used in inertial navigation, and  ring laser gyroscopes, used in geophysics\cite{chow85, vali76, stedman}.  {Furthermore, the Sagnac  effect due to the rotation of the Earth cannot be neglected in applications such as the Global Positioning System \cite{ashby-libro}, and  was also relevant for the famous experiment performed by Hafele and Keating in 1971 (and published in 1972) \cite{HK1, hk2, RS}}.

Today, the Sagnac effect is  important also for fundamental physics.  {In fact  large ring lasers can be used} for an independent determination of the Lense-Thirring effect, in an Earth based experiment \cite{StedLT,ginger11,ginger12}: these devices measure with great accuracy the rotation rate of the terrestrial laboratory where they are located with respect to an inertial frame, e.g. with respect to fixed stars and, in doing so, the gravitational drag of inertial frames comes into play (see for instance  \cite{MTW,ciufoliniwheeler}).  For practical motivations pertaining to the realization of such experiments, it is important to focus on the corrections to the leading term of the Sagnac effect  {which is proportional to the diurnal rotation rate $\Omega_{\oplus}$}.  For this reason, for instance, the effect of the deflection of light, due to the rotational inertial forces, on the Sagnac time delay was studied in \cite{maraner}. In this paper, we focus on the Sagnac effect  in flat space-time, to work out the impact of the higher order  kinematic corrections  {(proportional to $\Omega^{2}_{\oplus}$)  to the leading term, expressed by formula (\ref{eq:sagnac2}). The smallness of the General Relativistic effects sought for is such that the mentioned higher order terms could be comparable with the Lense-Thirring signal, masking or mimicking it. Elsewhere \cite{ginger11} we focused on the role of the gravitational field; here we are interested in the additional SRT small corrections.}

We will discuss to what extent eq. (\ref{eq:sagnac1})  is independent from the position of the center of rotation and from the shape of the enclosed area for an Earth-bound interferometer and, also, its relation with the Aharonov-Bohm effect.

The paper is organized as follows: in section \ref{sec:setup} we introduce the mathematical framework for describing the Sagnac effect, while in section \ref{sec:local} we apply the formalism to the case of a terrestrial laboratory. Conclusions are eventually drawn in section \ref{sec:conc}. We use the following notation: Greek and Latin indices
denote space-time and spatial components, respectively;
letters in boldface like $\mb x$ indicate spatial vectors. Moreover we use units such that $c=1$.

\section{Foundations of the Sagnac effect}\label{sec:setup}

In this Section we sketch the mathematical framework for the analysis of the Sagnac effect in a stationary space-time: a thorough approach can be found in \cite{kajari}.  To begin with,  both the observer (i.e. the measurement apparatus) and the experimental setup (i.e. the interferometer) are at rest in a reference frame which, in turn, is uniformly rotating with respect to an inertial frame. We choose a suitable set of  coordinates  $\{x^{\mu}\}=\{x^{0},x^{i}\}=\{x^{0},\mb x\}$\footnote{Greek indices run from 0 to 3, Latin indices from 1 to 3.} adapted to the rotating frame; because of stationarity, the space-time metric does not depend on time and it can be written in the form \footnote{We adopt the Einstein convention on the summation over repeated indices}
\beq
ds^{2}=g_{00}(\mb x)dt^{2}+2g_{0i}(\mb x)dtdx^{i}+g_{ij}(\mb x)dx^{i}dx^{j} \label{eq:metricastazionaria}
\eeq
 {The observer is located at a fixed spatial point $P$ in the rotating frame, whose position is  $\mb x_{P}$ and, from this point, two light beams are simultaneously emitted, and propagate in opposite directions along null curves that correspond to the spatial path of the interferometer. On setting $ds^{2}=0$, we are able to solve for the infinitesimal coordinate time interval along the world line of a light ray:
\beq
dt=\frac{-g_{0i}(\mb x) dx^{i} \pm \sqrt{g^{2}_{0i}(\mb x)(dx^{i})^{2}-g_{ij}(\mb x)g_{00}(\mb x)(dx^{i})(dx^{j})}}{g_{00}(\mb x)}
\label{eq:1rev}
\eeq
We are interested in solutions located in the future, so we choose $dt > 0$;  equation (\ref{eq:1rev}) allows to evaluate the coordinated time of flight of an electromagnetic signal between two successive events in vacuo. If we  consider a closed path (in space) and integrate
over the path  in two opposite directions from the emission to the absorption events, two different results for the times of flight are obtained because of the off diagonal $g_{0i}$ components of the metric tensor, say $t_{+}$, $t_{-}$, where ``$+$'' refers to the signal co-rotating with the reference frame, while ``$-$'' stands for the counter-rotating signal. Hence, we see that the difference between the times of flight turns out to be
\beq
\delta t= t_{+}-t_{-} = -2 \oint_{\ell} \frac{g_{0i}(\mb x)}{g_{00}(\mb x)}  dl^{i}
\eeq
where $\ell$ is the spatial trajectory of the beams, whose tangent vector is $d\mathbf l$; notice that, we have used the time independence of the metric coefficients, as well as the fact that emission and absorption happen at the same position in the rotating frame. Then, the observer located at $P$  measures  the proper-time difference:
\beq
\Delta \tau=-2 \sqrt{g_{00}(\mb x_{P})}  \oint_{\ell} \frac{g_{0i}(\mb x)}{g_{00}(\mb x)}  dl^{i} \label{eq:deltataulocal1}
\eeq}
We emphasized the spatial dependence of the metric elements in (\ref{eq:deltataulocal1}) because it is relevant for deriving the subsequent results.

The original Sagnac formula (\ref{eq:sagnac2}) is expressed in terms of area enclosed by the path of the beams: in order to see the connection with eq. (\ref{eq:deltataulocal1}), we need to use Stokes theorem. To this end, it is useful to define the fields
\beq
\mb h (\mb x) \doteq g_{0i}(\mb x), \quad \varphi(\mb x) \doteq \frac{1}{g_{00}(\mb x)} \label{eq:defvecscal1}
\eeq
In particular $\mb h(\mb x)$ and $\varphi(\mb x)$ are a vector and a scalar with respect to the coordinate transformation $x'^{i}=x'^{i}(x^{i})$  internal to the reference frame. Moreover,  {when dealing with the gravitational field of rotating objects, in the gravitoelectromagnetic formalism (see e.g. \cite{gem1,gem2})  $\mb h(\mb x)$ is usually referred to as the \textit{gravitomagnetic} potential, which enables to formally introduce the \textit{gravitomagnetic field} $\mb b(\mb x) \doteq \bm \nabla \wedge \mb h(\mb x)$. Since in this paper we are concerned with kinematic effects only and neglect gravity, in what follows $\mb b(\mb x)$ is merely related to the rotation of the reference frame.}

Then, we may write the Sagnac time delay in the form
\beq
\Delta \tau=-2 \sqrt{\frac{1}{\varphi(\mb x_{P})}}  \oint_{\ell} \varphi \mb h  \cdot d \mb l \label{eq:deltataulocal2}
\eeq
Now, by applying the Stokes theorem we may write the integral in (\ref{eq:deltataulocal2}) in the form
\beq
\oint_{\ell} \varphi \mb h  \cdot d \mb l = \int_{S} \left[\bm \nabla \wedge (\varphi \mb h) \right] \cdot d \mb S \label{eq:stokes1}
\eeq
where   $\mathbf S$ is the area vector of the surface enclosed by the contour line $\ell$. By using vector identities  and identifying the curl of $\mb h (\mb x)$ with $\mb b (\mb x)$ we obtain
\beq
\Delta \tau=-2 \sqrt{\frac{1}{\varphi (\mb x_{P})}} \int _{S} \left[\bm  \nabla \varphi(\mb x) \wedge \mb h  (\mb x)\right] \cdot d \mb S -2 \sqrt{\frac{1}{\varphi (\mb x_{P})}}  \int _{S} \left[\varphi (\mb x)  \mb b (\mb x) \right] \cdot d \mb S \label{eq:sagnac2area}
\eeq
This is the general expression of the Sagnac effect in terms of surface integrals. It is often said that the Sagnac effect can be expressed in analogy with the Aharonov-Bohm effect (see \cite{rizzi03a} and references therein): in other words the Sagnac effect is described in terms of the flux of the  field  {$\mathbf b(\mb x)$} across the interferometer area. By inspection, we see that if $\varphi(\mb x)$ is constant over $S$, or its change is negligibly small, the first integral in (\ref{eq:sagnac2area}) goes to zero, and the second is proportional to $\displaystyle \int_{S} \mb b(\mb x) \cdot d \mb S$.
As correctly reported by \cite{maraner} (see also \cite{RRinRRF}), this is just a formal analogy, because  in the case of the Aharonov-Bohm effect the {magnetic field} is null along the trajectories of the particles, while in the case of the Sagnac effect, the field  {$\mathbf b(\mb x)$}, which is proportional to the rotation rate vector (see below), is not null. Moreover, we have just seen that the exact expression (\ref{eq:sagnac2area}) can be written in terms of the flux of the field  {$\mathbf b(\mb x)$} across the interferometer area only  {if $\varphi(\mb x)$ is constant over $S$, or its change is negligibly small with respect to the main contribution:} this fact clearly limits the analogy between the two effects. In any case, in the following section we will give numerical estimates in order to evaluate to what extent the analogy holds true in actual terrestrial experiments.

\section{The Sagnac Effect in the Laboratory Frame}\label{sec:local}

Because of the actual interest in experiments exploiting Sagnac-based devices, such as ring lasers, it is important to evaluate the Sagnac effect in a terrestrial laboratory. As described in \cite{ginger11}, this can be done by defining the space-time metric in the ``proper reference frame'', that is to say nearby the world-line of the observer which performs measurements with the ring laser and moves in a given background reference frame. Indeed, here we are not interested in the effects due to the gravitational field of the Earth, rather,   we want to investigate the relativistic kinematic effects, due to the diurnal rotation. In order to define the suitable framework in which experiments are performed, we suppose that the observer carries an orthonormal tetrad and, to this tetrad, we associate the set of space coordinates $x^{i}$ and the observer's proper time $x^{0}$; in terms of these coordinates the metric coefficients in a neighborhood of the observer's world-line can be written in the form
(see e.g. \cite{MTW})
\beq
g_{00}=1+2 \bm{\mathcal{A}} \cdot \mb x, \quad  g_{0i}=\left(\bm{\Omega } \wedge \mb x \right)_{i}, \quad g_{ij}=-\delta_{ij} \label{eq:localmetric1}
\eeq
We emphasize that the metric (\ref{eq:localmetric1}) holds only near the world-line of the observer,
when  {corrections quadratically depending on the displacement are negligible}. In the above equations, $\bm{\mathcal{A}}$ is the spatial projection of the observer's four-acceleration (in the background reference frame), $\delta_{ij}$ are the spatial components of  the Minkowski tensor  {(when Cartesian coordinates are used)},  while the rotation rate vector $\bm \Omega$ is related to the parallel
transport of the basis four-vectors along the observer's world-line\footnote{In particular, if $\bm \Omega$ were zero,
the tetrad would be Fermi-Walker transported.}.  It is possible to show (see e.g. \cite{ginger11}) that, starting from a suitable metric which describes  the gravitational field of the
rotating Earth, and taking into account the motion of the Earth-bound laboratory, the rotation rate $\bm \Omega$ turns out to be $\displaystyle \bm \Omega= -\bm \Omega_{0}-\bm \Omega_{T}-\bm \Omega_{G}-\bm \Omega_{B}-\bm \Omega_{W}$,
where $\bm \Omega_{0}$ is the Earth rotation rate, as measured in the local frame,  $\bm \Omega_{T}$ is the Thomas precession, $\bm \Omega_{G}$ is the geodetic or de Sitter precession, $\bm\Omega_{B}$ is the Lense-Thirring precession, and  $\bm \Omega_{W}$ is due to the preferred frame effect. Since here we are interested in the kinematic effects only, we disregard the contributions of gravitational origin $\bm \Omega_{G}, \bm \Omega_{B}, \bm \Omega_{W}$ due to the mass and angular momentum of the Earth, and we obtain
\beq
\bm \Omega =  -\bm \Omega_{0}-\bm \Omega_{T} \label{eq:Omegatot2}
\eeq
 { Given an Earth-bound laboratory moving at the speed $V= \Omega _{\oplus} R_{\oplus} \sin \theta$ with respect to an inertial frame, where $R_{\oplus}$ is the terrestrial radius,  $\theta$ is the colatitude angle of the laboratory location and $\bm \Omega_{\oplus}$ is the terrestrial rotation rate, as measured in an asymptotically flat inertial frame, on applying Lorentz time dilation, we get for  the Earth rotation rate, as measured in the laboratory frame, the expression:}
$\bm \Omega_{0} \simeq \left[1+\frac{1}{2}\Omega^{2}_{\oplus} R^{2}_{\oplus}  \sin^{2} \theta \right]\bm \Omega_{\oplus}$.
 As for the Thomas precession its expression turns out to be  {(see e.g. \cite{ginger11})} $\displaystyle \bm \Omega_{T}=-\frac{1}{2} \Omega^{2}_{\oplus} R_{\oplus}^{2} \sin^{2} \vartheta \, \bm{\Omega}_{\oplus}$. Hence to lowest order in $\Omega_{\oplus} R_{\oplus}$, we get
\beq
\bm \Omega = -\left[1+ \frac{1}{2}\Omega^{2}_{\oplus} R_{\oplus}^{2}  \sin^{2} \theta \right]\bm \Omega_{\oplus}+\frac{1}{2} \Omega^{2}_{\oplus} R_{\oplus}^{2} \sin^{2} \vartheta \bm{\Omega}_{\oplus}= -\bm \Omega_{\oplus} \label{eq:Omegalocal00}
\eeq
Moreover, if we neglect gravitational terms, for an Earth-bound laboratory, on using cylindrical base vectors $\mb u_{\rho}, \mb u_{\vartheta}, \mb u_{z}$, it is possible to write the laboratory acceleration  $\bm{\mathcal{A}} $  {with respect to a reference frame at rest with respect the center of the  Earth (e.g. the International Terrestrial Reference System, ITRF) and the rotation rate (\ref{eq:Omegalocal00}) in the form}
\beq
\bm{\mathcal{A}}= - \Omega^{2}_{\oplus} \, R_{\oplus} \, \sin \theta \, \mb u_{\rho},  \quad \bm{\Omega}= - \Omega_{\oplus}  \, \mb u_{z} \label{eq:AOmegalocal}
\eeq
Now,  it is possible to apply our general relation (\ref{eq:sagnac2area}). In this case, it is
\beq
\varphi(x^{i})=\frac{1}{1+2\bm{\mathcal A} \cdot \mb x}, \quad \mb h(x^{i})= \left(\bm{\Omega } \wedge \mb x \right) \label{eq:defphhlab}
\eeq
In particular, we see that   $\mb b= 2 \bm \Omega$.  On substituting in (\ref{eq:sagnac2area}) we obtain
\beq
\Delta \tau=-2 \sqrt{1+2\bm{\mathcal A} \cdot \mb x_{P}} \int _{S} \left[\frac{-2 \bm{\mathcal A} \wedge  \left(\bm{\Omega } \wedge \mb x \right) }{\left(1+2\bm{\mathcal A} \cdot \mb x \right)^{2}} \right] \cdot d \mb S -2 \sqrt{1+2\bm{\mathcal A} \cdot \mb x_{P}}  \int _{S} \left[\frac{\bm \nabla \wedge \left(\bm{\Omega } \wedge \mb x \right) }{1+2\bm{\mathcal A} \cdot \mb x} \right] \cdot d \mb S \label{eq:sagnaclocal11}
\eeq
and then, by performing the vector operations
\beq
\Delta \tau=4 \sqrt{1+2\bm{\mathcal A} \cdot \mb x_{P}} \int _{S} \left[\frac{\bm \Omega \left( \bm{\mathcal A} \cdot \mb x \right)- \mb x \left(\bm{\mathcal A} \cdot \bm \Omega \right)}{\left(1+2\bm{\mathcal A} \cdot \mb x \right)^{2}} \right] \cdot d \mb S -4 \sqrt{1+2\bm{\mathcal A} \cdot \mb x_{P}}  \int _{S} \left[\frac{\bm \Omega }{1+2\bm{\mathcal A} \cdot \mb x} \right] \cdot d \mb S \label{eq:sagnaclocal22}
\eeq
Since, in this case, $\bm{\mathcal A} \cdot \bm \Omega=0$, choosing the origin in correspondence of the observer (i.e. $\mb{x}_P=0$)  {and taking into account that we neglect corrections quadratically depending on the displacements}, we may write:
\beq
\Delta \tau=4  \int _{S} \left[{\bm \Omega \left( \bm{\mathcal A} \cdot \mb x \right)} \right] \cdot d \mb S -4  \int _{S} \left[\frac{\bm \Omega }{1+2\bm{\mathcal A} \cdot \mb x} \right] \cdot d \mb S \label{eq:sagnaclocal23}
\eeq
We see that the Sagnac effect depends, in general, both (i) on the position of the interferometer in the rotating frame through the acceleration $\bm{\mathcal A}$, whose expression is related to the laboratory location on the Earth, and (ii) on the interferometer size, since the integrands in (\ref{eq:sagnaclocal23}) are not constant across the interferometer area.

In order to evaluate the impact of these effects, it is useful to introduce the dimensionless parameter {\footnote{Remember that $c=1$ in our units.}}
\beq
\varepsilon \doteq \bm{\mathcal A} \cdot \mb x \simeq {\Omega_{\oplus}^{2} R_{\oplus}L}{}  \simeq 4 \times 10^{-19} \left(\frac{L}{\mathrm{1 \ m}} \right) \label{eq:localorder1}
\eeq
 where $L$ is the linear size of the interferometer.  As a consequence, to zeroth order in $\varepsilon$, we get
\beq
\Delta \tau = \Delta \tau_{0}= -4  \int _{S} \bm \Omega \cdot d \mb S= 4  \int _{S} \bm \Omega_{\oplus} \cdot d \mb S = 4 \bm \Omega_{\oplus} \cdot \mb S \label{eq:sagnaclocal0}
\eeq
that is the original Sagnac formula (with $c=1$), as expected, and the analogy with the Aharonov-Bohm effect holds true. A development in power series of $\varepsilon$ allows to evaluate the Sagnac effect up to all required orders: however, it is manifest that the corrections to $\Delta \tau_{0}$ deriving both  from the first and the second integral in (\ref{eq:sagnaclocal23}) are proportional to $\varepsilon$.  In terms of rotation rate measured by ring lasers, this corresponds to a relative sensitivity that is at least $\Delta \Omega /\Omega \simeq 10^{-19}$.  This is some 11 orders of magnitude smaller than the relative sensitivity currently available from the ring laser  in Wettzell \cite{laserG}, and 8 orders of magnitude smaller than the expected sensitivity of the GINGER experiment, aimed at measuring the Lense-Thirring effect \cite{ginger11,ginger12}.

\section{Conclusions}\label{sec:conc}

We studied the Sagnac effect for light beams in flat space-time, in order to evaluate the possible higher order corrections of kinematic origin to the Sagnac formula (\ref{eq:sagnac2}). In particular, we focused on the relevance of these terms for terrestrial experiments that are now being planned. To this end, we worked out the necessary formalism in the context of the local laboratory frame and, in order to make a connection with the Aharonov-Bohm effect, we derived an exact expression of the Sagnac effect in terms of surface integrals, across the interferometer area. We showed that the analogy with the Aharonov-Bohm effect holds true to lowest approximation order only and that,  in general, the Sagnac effect is influenced by both  the position of the interferometer in the rotating frame and  its extension. The expressions that we obtained can be developed in power series of a suitable dimensionless parameter, to obtain all kinematic corrections. However, as for the accuracy available for the experiments that are now under investigation, the lowest order approximation is sufficient, so that the special relativistic or kinematic Sagnac effect is consistently described by the expression (\ref{eq:sagnac2}).

\end{document}